\documentclass[12pt,dvips]{article}
\usepackage{epsfig}

\begin{document}
%%%%%%%%%%%%%%%%%%%%%%%%%%%%%%%%%%%%%%%%%%%%%%%%%%%%%%%%%%%%%%%%%%%%%%%%%%%

\thispagestyle{empty}

\date{July 3, 1998}
\title{
\vspace{-5cm}
\begin{flushright}
{\normalsize UNIGRAZ-}\\
\vspace{-4mm}
{\normalsize UTP-}\\
\vspace{-4mm }
{\normalsize 03-07-98}\\
\end{flushright}
\vspace*{3cm}
The lattice Schwinger model\\with the SW action}

\author{Ch.~Hoelbling$^a$\footnote{E-mail: hch@bu.edu}, \\ 
C.~B. Lang$^b$
and R.~Teppner$^b$\\~\\
$^a$ 
\normalsize Department of Physics, 
\normalsize Boston University\\
\normalsize Boston, MA 02215,
USA\\
\\
$^b$ 
\normalsize Institut f\"ur Theoretische Physik,\\
\normalsize Universit\"at Graz, A-8010 Graz, AUSTRIA \\
\\
}
\maketitle

\begin{abstract}
We perform a model study on the $2$-flavor lattice Schwinger model
using standard Wilson and ${\cal O}(a)$ improved Sheikholeslami-Wohlert
(SW) action. We find, that the phase diagram is altered, the critical line
shifted closer towards its continuum value $\kappa_c=0.25$. We find no
improvement in the rotation invariance of meson propagators; the
scaling of the Schwinger mass is considerably improved, high momentum
states are not. The additional cost of $\approx 30 \%$ CPU-time is
highly justified when calculating masses, but not for high momentum
observables.
\end{abstract}

\noindent


\newpage 
\section{Introduction}
In recent years, there has been an increasing interest in possible
use of improved actions to lattice QCD. There exist different
improvement schemes for lattice actions and observables, all which aim
at the extraction of continuum physics from simulations on coarse
lattices.

The simplest of these improvement schemes is the clover action, an 
${\cal O}(a)$ improvement of
the Wilson fermion action suggested by Sheikholeslami and Wohlert
\cite{ShWo85}, who implemented the Symanzik improvement \cite{Sy83} for
the Dirac operator. Several calculations in quenched and some in full
QCD have been performed using this action especially
to calculate both heavy and light hadron masses. The results obtained
seem promising, so that one might want to check the improvement
effects of the clover action in a simple and well-known toy model.
Here we study the effects of the ${\cal O}(a)$ improvement of the Wilson
action in the full lattice Schwinger model. This model
seems optimally suited for this purpose. 

Massless QED in $1+1$ dimension was first studied by Schwinger
\cite{Sc62b} as an example of an explicitly solvable QFT. In
subsequent papers \cite{KoSu} the model was generalized to allow
for fermion masses and include different flavors.

The Schwinger model became of considerable interest during the
early 1970's, because it shows some remarkable features -- like fermion
confinement or a $\Theta$-vacuum structure -- which are reminiscent of
QCD. For the purpose of this paper, it is sufficient to note, that the
massless $N$-flavor Schwinger model has a spectrum which consists
of one massive isosinglet boson and a isomultiplet of $(N^2-1)$
massless fermion-antifermion bound states
(for further discussion cf. e.g. \cite{GaSe94}). The mass of the
isosinglet state, the 'Schwinger mass' (in lattice units) is given by
\begin{equation}
m_s=\sqrt{\frac{N}{\beta\,\pi}} \quad .
\end{equation}
where $\beta=1/(e^2a^2)$ denotes the dimensionless gauge coupling.

In the next section some technical details about the definition and
the numerical simulation will be given and in Sec. 3 we discuss the
effect of the improvement on the rotation invariance of the propagator,
the phase diagram and the dispersion relations of the Schwinger boson.

\section{${\cal O}(a)$ improvement and simulation}

The fermionic part of the Wilson action for the massive, N-flavor
Schwinger  model is
\begin{equation}
 S_W=\sum_x \bar{\Psi}_x\Psi_x-\kappa\sum_{x,\mu} 
  \bar{\Psi}_{x+\hat{\mu}}(1+\sigma_\mu)\,U^{\dag}_\mu\Psi_x +
  \bar{\Psi}_x(1-\sigma_\mu)\,U_\mu\Psi_{x+\hat{\mu}} \quad .
\end{equation}
The fields have $2N$ components, $\sigma_\mu$ are the Pauli matrices.
This action has a classical ${\cal O}(a)$ error introduced by the
Wilson term, which can be removed \cite{ShWo85} by rotating the fermion
fields
\begin{equation}
\label{frot}
\Psi\rightarrow(1-\frac{1}{2}\,{\slash\!\!\!\!D})\Psi\quad ,\qquad
\bar{\Psi}\rightarrow\bar{\Psi}(1+\frac{1}{2}\,\stackrel{\leftarrow}
{\slash\!\!\!\!D}) 
\end{equation}
and adding a term
\begin{equation}
 S_{SW}=-\kappa c_{SW}\frac{i}{2}\,\sum_x F_{\mu\nu}
 \sigma_{\mu\nu}\bar{\Psi}_x\Psi_x
\end{equation}
to the action. $F_{\mu\nu}$ denotes the clover-leaf
discretization of the field strength tensor.
The resulting lattice Dirac operator now has 
discretization errors of ${\cal O}(a^2)$.

We note in passing that for this model one may introduce a
topological charge in its `geometric' definition
$\nu\propto\sum_{\mu,\nu} F_{\mu,\nu}$ \cite{TopDef}. Thus $S_{SW}$
introduces a local coupling of $\bar{\Psi}\Psi$ to $\nu\,\sigma_3$.

The coefficient $c_{SW}$, which generally has to be determined
non-per\-turba\-tive\-ly at finite $\beta$, is just $c_{SW}=1$ in 2D, since
the gauge coupling $e$ is dimensionful and therefore $c_{SW}=1+{\cal
O}(e^2\,a^2)$, which only gives corrections to higher order in $a$.

Since the gauge part of the Wilson action
\begin{equation}
S_{G}=\frac{1}{2}\sum_p\mbox{Tr}(U_p+U^{\dag}_p)
\end{equation}
  has ${\cal O}(a^2)$ discretization errors, the total action
\begin{equation}
 S=S_W+S_{SW}+S_{G}
\end{equation}
together with (\ref{frot}) gives a lattice version of the Schwinger
model with discretization errors of ${\cal O}(a^2)$.

We have done a MC simulation of the Schwinger model, using both the
Wilson and the SW action, with two degenerate flavors of dynamical
fermions. We used a hybrid Monte Carlo algorithm with 10 integration
steps and a trajectory length tuned for a $0.8$ acceptance rate of the
MC step. For fermion matrix inversion we used a standard conjugate
gradient algorithm, and there was no problem with its convergence or
stability. In that algorithm the doubling of the fermionic species  is
required in order to guarantee positivity of the fermionic determinant 
-- one of our reasons to study the Schwinger model with two flavors.

In our implementation, the speed loss when including the clover term
into the action is consistently $\approx 30\%$.

Depending on the lattice size (from $8\times 8$ up to $18\times 18$)
and couplings, we skipped $20$ to $100$ configurations between
measurements and finally had between $2\times10^3$ and $10^4$ independent
configurations. To check for ergodicity, we measured the topological
charge (using the geometric definition) and observed,
that in each run the system tunneled several times between different
topological sectors.

\section{Results}

\subsection{Phase diagram}

The additional term in the action affects the critical $\kappa$ value
where the theory has effectively massless fermions. Since the clover
action is supposedly closer to the continuum (has smaller corrections
to leading scaling behavior), we might expect less renormalization
of $\kappa_c$ with a value closer to its continuum value of $0.25$ even
for small $\beta$.

\begin{figure}
\begin{center}
\epsfig{file=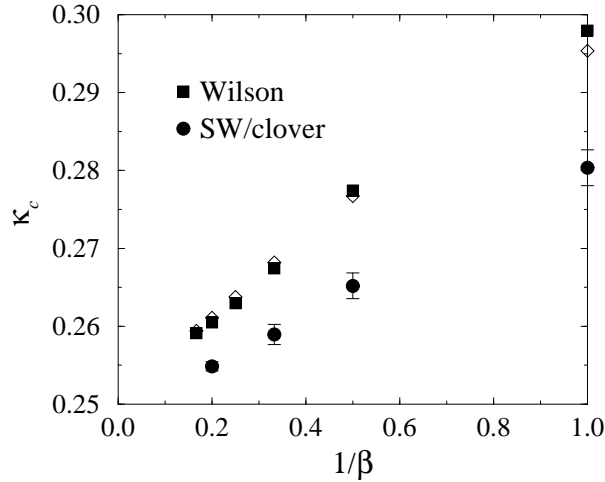,width=8cm}
\end{center}
\caption{\label{phd}Phase diagram for Wilson and clover action on an
$8\times 8$ lattice. The diamonds indicates the 
results (Wilson action) of an extrapolation to infinite lattice 
volume from results at various lattice sizes \cite{HiLaTe98}. 
We find that $\kappa_c$ is closer to its continuum value 0.25 
for the clover action.}
\end{figure}

In order to determine $\kappa_c(\beta)$ we follow suggestions
\cite{PCAC} to utilize the PCAC relation as discussed in
\cite{HiLaTe98} for the Schwinger model.  At  each $\beta$ we measured
an observable proportional to the effective fermion mass for $5$
different $\kappa$ values and determined $\kappa_c$. Our result for a
$8\times 8$ lattice is compared with the result for the original Wilson
fermion action in Fig. \ref{phd}.

The values $\kappa_c$ for the clover action are in fact closer to
$0.25$. However, both, Wilson and clover action have corrections
\begin{equation}
\kappa_c=0.25+{\cal O}(a^2)=0.25+ {\cal O}(1/\beta) \;.
\end{equation}
The dependence on the finite lattice volume is weak and -- for
the Wilson action -- is discussed elsewhere \cite{HiLaTe98}.

\subsection{Spectrum of the Dirac operator}

\begin{figure}
\begin{center}
\epsfig{file=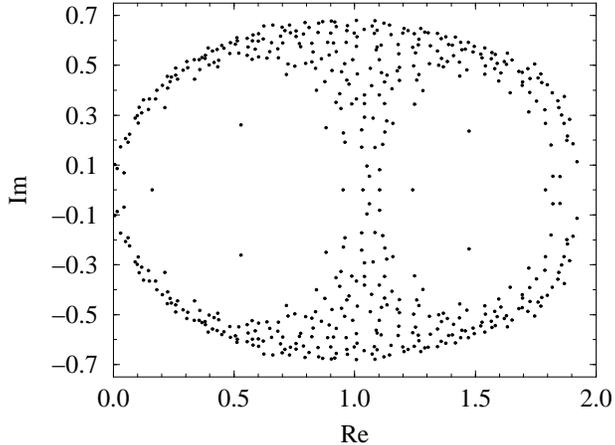,width=8cm}
\end{center}
\caption{\label{SWspectrum}The typical shape of the eigenvalue spectrum for the SW
action; here shown for a configuration on a $16\times 16$ lattice
determined at $\beta=2$ and
$\kappa_c$.}
\end{figure}

The spectral distribution of the Dirac operator is of some interest with
regard to topological and chiral properties of the system. Recent
studies for the Schwinger model 
\cite{NaNeVr95,BaDuEi98b,GaHiLa97a,FaLaWo98} and for the SW-action in 4D
\cite{SW4D} have emphasized these aspects. Fig.
\ref{SWspectrum} should be compared with typical spectra for the pure
Wilson Dirac operator (cf. e.g. \cite{GaHiLa97a}). 

As compared to the Wilson action  with its
$\lambda\leftrightarrow\bar\lambda$ and
$(1-\lambda)\leftrightarrow(\lambda-1)$ symmetries we observe for the
SW operator an agglomeration of eigenvalues at larger distances from 0.
This feature may improve somewhat the separation of the low-lying
states from the doubling states already at moderate $\beta$. For the
continuum limit the distribution density of small eigenvalues is of
relevance. However, the distribution is not much closer to the circular
shape one obtains for fixed point actions \cite{HaSpec} or
Neuberger-projected actions \cite{NeProj}.  We cannot identify a clear
signal of improvement in this respect for the SW Dirac operator.

\subsection{Rotation invariance}

We measured the propagators of the pseudoscalar isotriplet ($\pi$)
mesonic bound states. In the 2-flavor model these
modes are expected to be massless at $\kappa_c$. For short distances on
the lattice, the rotation invariance of these propagators is clearly
broken when using Wilson action. Generally, improved actions are
expected to show better rotation invariance (cf. the Schwinger model FP
action study in \cite{LaPa98b}). 

One must be careful not to compare these quantities simply at the same
values of $\beta$ and $\kappa$, but at values corresponding to
comparable discretization of the same continuum theory. We choose to
compare the propagators at $\beta=2$ and the respective $\kappa_c$, so
that we have a discretization of the massless (fermion) theory in both
cases.

\begin{figure}
\begin{center}
\epsfig{file=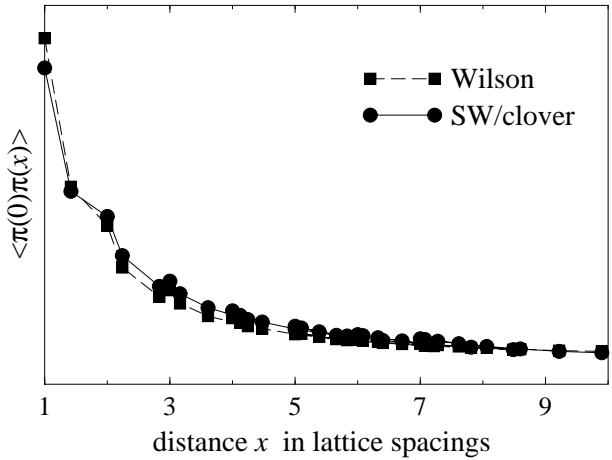,width=8cm}
\end{center}
\caption{\label{rot}Plot of $\pi$ propagator
as measured at $\beta=2$ on a $16\times 16$ lattice at the
respective $\kappa_c$. Using the clover action gives no improvement in
the rotation invariance of the propagator.}
\end{figure}

The results are shown in Fig.\ref{rot}. We find no noticeable
improvement of rotation invariance when using the clover action. The
relative error of the diagonal $(1,1)$ propagator compared to the
$(1,0)$ and $(2,0)$ propagators is $\approx (18\pm 1)\%$ for both
actions. This is the first indication, that short distance (high
momentum) observables are not improved.

\subsection{Meson dispersion relations}

\begin{figure}
\begin{center}
\epsfig{file=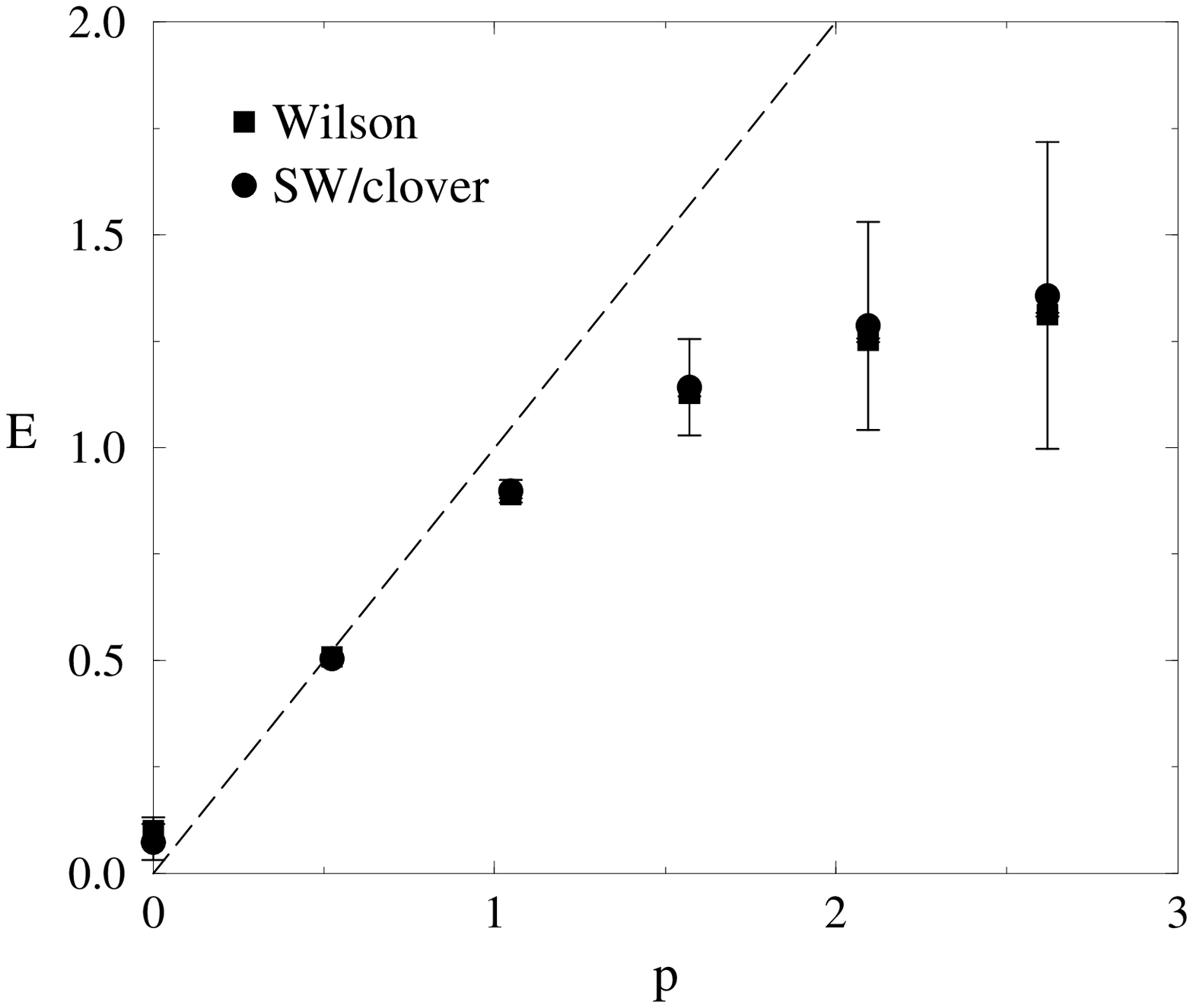,width=6.5cm}\hfil
\epsfig{file=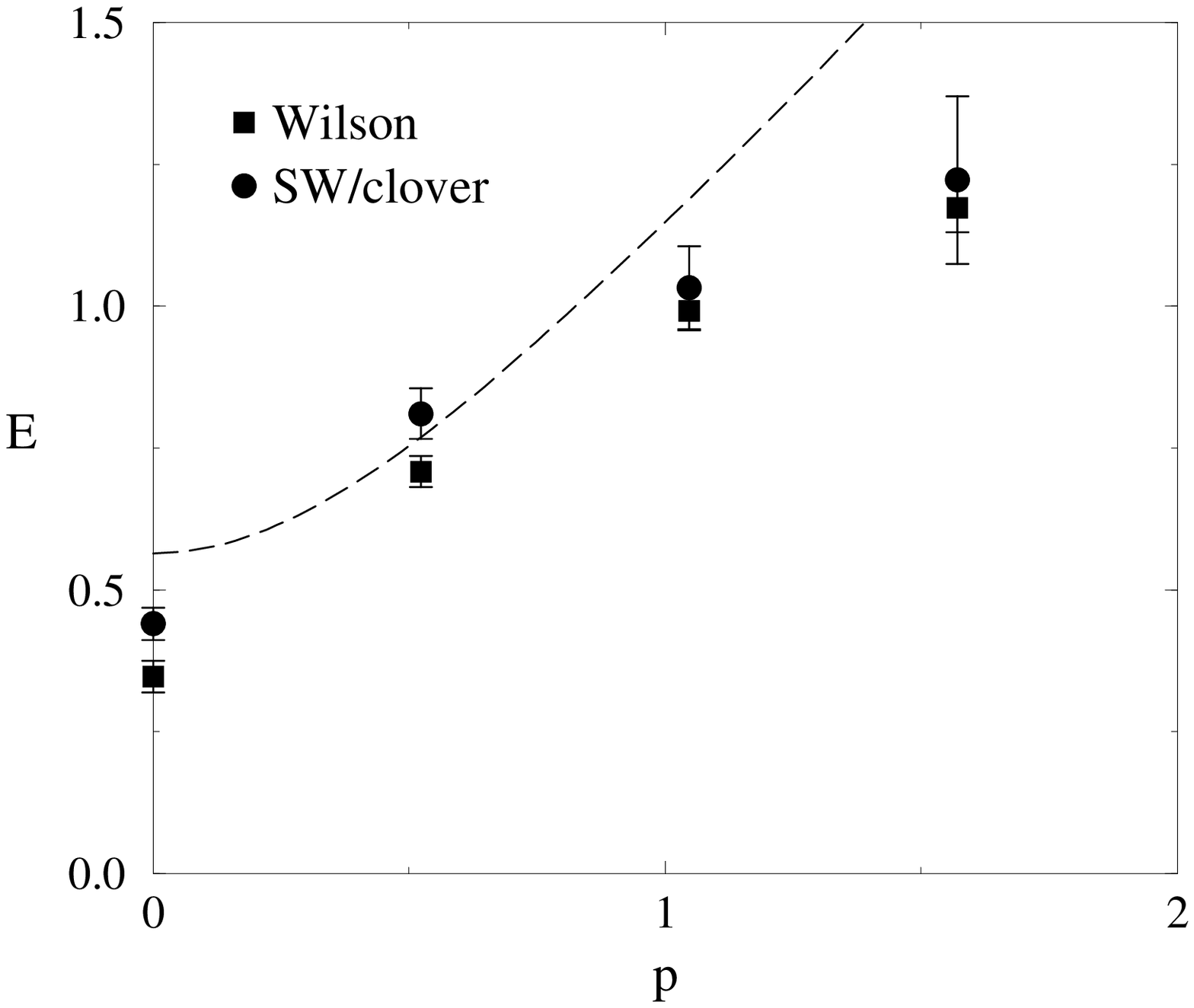,width=6.5cm}
\end{center}
\caption{\label{disp}Dispersion relation (for $\beta=2$, lattice size $12\times 12$)
for a massless (isotriplet-vector) and the massive (isosinglet-vector) 
meson. The dashed curves denote the continuum dispersion relations.
High momentum values are not improved by the clover action.}
\end{figure}

Fig. \ref{disp} shows the dispersion relations for the massive
(isosinglet vector) and the massless (isotriplet vector) meson at 
$\beta=2$ and
$\kappa=\kappa_c$ on a $12\times 12$ lattice.  No improvement for the
high momentum behavior is observed.

\begin{figure}
\begin{center}
\epsfig{file=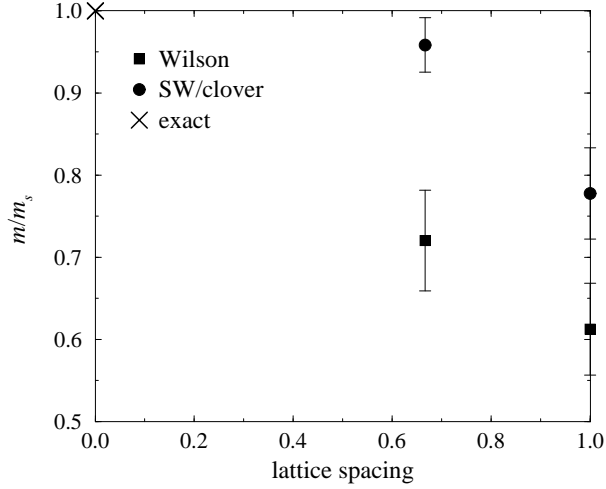,width=8cm}
\end{center}
\caption{\label{mass}The Schwinger mass as measured at $\beta=2$ on a
$12\times 12$ and $\beta=4.5$ on a $18\times 18$ lattice over the
continuum Schwinger mass. For this low momentum observable, the
improvement is significant. The lattice spacing is given in
multiples of the value at $\beta=2$, i.e. $\sqrt{2/\beta}$.}
\end{figure}

Concerning the mass of the massive boson, however, the improvement is
significant. Fig. \ref{mass} shows the ratio of the observed Schwinger
mass over its continuum value at $\beta=2,\kappa=\kappa_c$ on a
$12\times 12$ and $\beta=4.5$, $\kappa=\kappa_c$ on a $18\times 18$
lattice.
Assuming the continuum scaling behavior $a(\beta)\propto
1/\sqrt{\beta}$ these are different discretizations of the same physical
system size $L\,a\propto L/\sqrt{\beta}\simeq 8.5$. 
The improvement when including the clover term is quite
obvious. The scaling corrections are remarkably smaller for the clover
results.

This behavior, the considerable improvement of low momentum states and
the lack thereof for high momentum states, is no surprise when one
considers, that the construction of Symanzik improved actions aims at
improving the fermion dispersion relations for low momenta (cf.
\cite{ShWo85}). The overall improvement of the fixed point action in
\cite{LaPa98b} (which, however has more than hundred terms per site in the
action) is definitely more pronounced and extends to high momenta as
well.

\section{Conclusion}

We performed a Monte Carlo simulation on the 2-flavor lattice Schwinger
model, comparing standard Wilson to an ${\cal O}(a)$ improved
Sheikholeslami-Wohlert action. We found the following results:

\begin{itemize}
\item
The critical value of the hopping parameter $\kappa_c$ 
moves closer to its continuum value $0.25$.
\item
The eigenvalue spectrum of the lattice SW Dirac operator changes but
does not appear to improve significantly towards, e.g. a circular shape.
\item
There was no improvement in the rotation invariance of meson propagators.
For the meson dispersion relations, there is no improvement in the 
high momentum states.
\item
There is considerable improvement in the scaling behavior of the
Schwin\-ger mass of the massive boson.
\end{itemize}

These observations are consistent with the fact that the
SW action is constructed as an improved action for
low momentum fermionic states $p\ll 1/a$. While there is considerable
improvement in low momentum observables, high momentum observables are
largely unaffected by the addition of the clover term.

For the full fermion QCD simulations this suggests that it is
preferable to use the clover action as long as one is only interested in
the mass spectrum of the theory. On the other hand, some observables
will likely show no improvement at all, especially those connected with
high momentum states.

\subsection*{Acknowledgment} 

We would like to thank Prof. Claudio Rebbi and Ivan Hip for many
helpful discussions. C.B.L. also wants to thank Claudio for the kind
hospitality at Boston University and wants to acknowledge support by
Heinrich-J\"org-Stiftung, Univ.  Graz. The work was supported by Fonds
zur F\"orderung der Wissenschaftlichen Forschung  in \"Osterreich,
Project P11502-PHY.

%\bibliographystyle{npb}
%\bibliography{/u1/people/cbl/refs/lgt+1990,/u1/people/cbl/refs/lgt-1989}

\end{document}